\documentclass{mem}
\usepackage{natbib}
\usepackage{txfonts}
\usepackage{balance}
\usepackage{graphicx}
\usepackage[a4paper]{hyperref}
\idline{75}{282}
\begin{document}
\def\teff{$T\rm_{eff }$}
\def\kms{$\mathrm {km s}^{-1}$}

\title{The Chemical Composition of Cepheids in the Milky Way and the Magellanic Clouds}

%   \subtitle{}

\author{M. Mottini \inst{1}
          \and M. Romaniello \inst{1}
	  \and F. Primas \inst{1}
          \and G. Bono \inst{2}
          \and M.A.T. Groenewegen \inst{3}
          \and P. Fran\c{c}ois \inst{4}
           }

  \offprints{M. Mottini}
 
\institute{European Southern Observatory, Karl-Schwarzschild-Strasse 2, D-85748 Garching bei M{\"{u}}nchen, Germany
          \and INAF-Osservatorio Astronomico di Roma, via Frascati 33, I-00040 Monte Porzio Catone, Italy
          \and Instituut voor Sterrenkunde, Celestijnenlaan 200B, B-3001 Leuven, Belgium
           \and Observatoire de Paris-Meudon, GEPI, 61 avenue de l'Observatoire, F-75014 Paris, France}

\authorrunning{Mottini et al.}

\titlerunning{Chemical Composition of Cepheids}

\abstract{We have measured the elemental abundances of 68 Galactic and Magellanic Cepheids from FEROS and UVES high-resolution and high signal-to-noise spectra in order to establish the influence of the chemical composition on the properties of these stars \citep[see][]{paper1}. Here we describe the robust analytical procedure we have developed to accurately determine them. The resulting iron abundances span a range between $\sim -$0.80 dex for stars in the Small Magellanic Cloud and $\sim$ +0.20 dex for the most metal-rich ones in the Galaxy. While the average values for each galaxy are in good agreement with non-pulsating stars of similar age, Cepheids display a significant spread. Thus it is fundamental to measure the metallicity of individual stars.

\keywords{Stars: abundances --
                Cepheids}
}
\maketitle{}

\section{Introduction}

Cepheid stars, through their Period-Luminosity (PL) relation, are one of the key tools in determining the extragalactic distance scale. In spite of its fundamental importance, to this day, it is still unclear the role played by the chemical composition on the pulsational properties of these stars.
Recent theoretical works give different results: some authors claim that metallicity has negligible effects on the PL relation \citep{chiosi92,san99,bar01}; others claim that there is a significant dependence of the PL relation on metallicity \citep{bono99,fio02,mar05}.

Although most of the observational efforts aiming at deriving the chemical composition of Cepheids and its effects on PL relation have used indirect means, like the measurement of oxygen abundances in the \ion{H}{ii} regions \citep[e.g.][]{sas97,ken98,sak04}, some of them have focused on a direct chemical analysis \citep{fc97,lu98,and02,lu03}.
Our study follows this second approach: we analysed a sample of high resolution spectra and derived $\alpha$-elements and iron abundances spectroscopically. The preliminary results for a sub-sample of 37 stars have already been presented in \citet{paper1}. Here we present the complete sample.

\section{The data sample}

We used the following selection criteria to choose the Cepheids of this data set: stars from three galaxies (Milky Way, LMC and SMC) in order to cover a factor of about ten in metallicity, well-determined intensity-mean magnitude in B, V and K bands for all the Cepheids \citep{ls94,groe04,storm04}; a wide range of periods in order to cover the entire PL relation (from 2 to 98 days).

We have a sample of 68 Cepheids: 32 Galactic Cepheids, 22 LMC and 14 SMC Cepheids.
The spectra of the Galactic Cepheids have been collected with FEROS (R=48\,000, spectral coverage: 3700-8600~\AA) at the ESO 1.5 meter telescope (La Silla) with a $S/N\sim150$ for $\lambda > 5000$~\AA, $S/N \sim 70$ for $\lambda > 4000$~\AA.
The spectra of the Magellanic Cepheids have been collected with UVES (R=40\,000, spectral coverage: 4800-6800~\AA\ for the red arm) at the ESO VLT-Kueyen (Paranal) with a $S/N\sim70$ for $\lambda > 5800$~\AA, $S/N\sim50$ for $\lambda > 4800$~\AA.

\section{The analysis}

In order to derive elemental abundances one needs to know the stellar parameters of the star and have a reliable line list available.

As a first step we carefully assembled the line list. We have assembled our \ion{Fe}{i}-\ion{Fe}{ii} and $\alpha$-elements line list from \citet{clem95}, \citet{fc97}, \citet{kv00} and \citet{and02} including a selection of lines  from VALD \citep{kup99,ryab99}. The VALD lines have been selected for effective temperatures typical of Cepheid stars (4500-6500 K). We have, then, visually inspected each line on the observed spectra, in order to check their profile and to discard blended lines. In order to do so, we have searched the VALD database (with the command {\em extract stellar}) for all the listed lines between 4800 and 7900~\AA\ on stellar spectra characterised by parameters typical of Cepheids (\teff=4500~K, 5500~K and 6500~K, $\log g$=2 and v$_t$=3~\kms). We then overplotted all the lines found in VALD, that fall within $\pm$3 \AA\ from each of our lines, and checked for their possible contribution to the equivalent width of our line. Any contribution larger than 5\% of the line strength made us discarding the line under scrutiny.

Our final list includes 246 \ion{Fe}{i} lines and  17 \ion{Fe}{ii} lines spanning a spectral range from 4800 \AA\ to 7900 \AA\, which is the range covered by our FEROS spectra. For the UVES spectra this list is slightly reduced due to their narrower spectral range. Simultaneously we also looked at $\alpha$- elements (O, Na, Mg, Al, Si, Ca and Ti) and we have assembled a list of almost 200 lines.
 For all the lines we have adopted the physical properties (oscillator strengths, excitation potentials) listed in VALD.

As a second step we measured the equivalent widths (EW) of all the lines assembled as described above. Because of their large number, we have used a semi-interactive routine developed by one of us (PF, {\em fitline}). This code is based on genetic algorithms, which mimic how DNA can be affected to make the evolution of species \citep{char95} and it uses a Gaussian fit. For about 15\% of the lines, the Gaussian profile adopted by {\em fitline} could not satisfactorily reproduce the observed profile. For these cases (usually very broad or asymmetric lines), the equivalent widths were measured manually with the IRAF {\em splot} task. The mean difference, as computed for those lines for which both methods could be applied, is around 1.5 m\AA, comparable to the average error on the EW inferred from the quality of the data \citep[Equation 7,][]{cay88}. The outcome of this test is that we can then safely use all our EW measurements, independently of the method followed to derive them. 

With these EWs in hand we then derived the microturbulent velocity and gravity of the studied star. We used as first guess values typical of Cepheids as inferred from previous studies (see Section 1 for references) then we got the final constraints from the abundance analysis: from the minimization of the log(Fe/H) vs EW slope for the microturbulent velocity and from the ionization balance for the log g. For the determination of the effective temperature we used the line depth ratios method described in \citet{kg00}.

Finally we determined the elemental abundances of our stars by using the Kurucz WIDTH9 code \citep{ku93} and LTE model atmospheres computed without the overshooting approximation and with the most recent opacity distribution functions \citep{cas03}.

\section{Results and future developments}

\begin{figure*}[]
\includegraphics[width=12cm,clip=true]{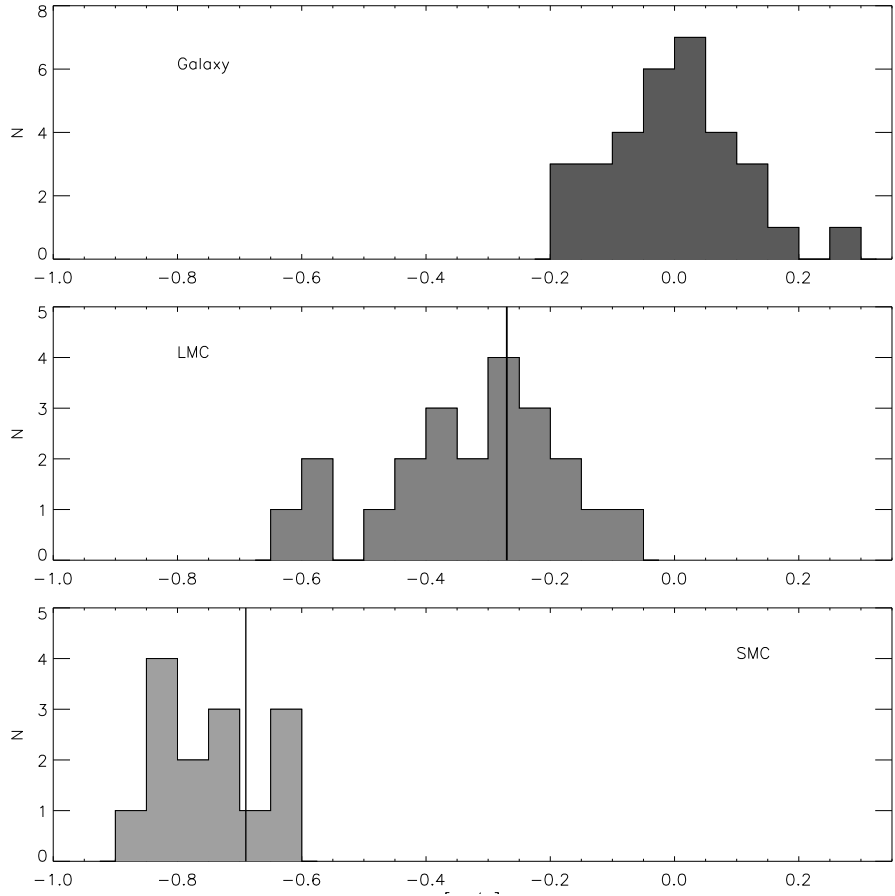}
\caption{\footnotesize Histograms of the [Fe/H] ratios derived for all the stars of our sample in the Galaxy, the LMC and the SMC. The solid lines indicate the mean values found using F and K supergiants by \citet{hill95} and \citet{hill97}.}
\label{abu-all}
\end{figure*}

Figure~\ref{abu-all} shows the distributions of the iron abundances of our programme stars in the Galaxy and the Magellanic Clouds. In the latter case we compare our results with the mean values obtained by \citet{hill95} and \citet{hill97} for F and K supergiants in the LMC and SMC, respectively. We note a good agreement with our mean values of the iron content.
For our Galactic sample, we find that the mean value of the iron content is solar ($\sigma$ = 0.10), with a range of values between $-$0.18~dex and +0.25~dex. These two extremes are represented respectively by 2 and 1 stars (out of the 32 in total). We note that more than 50\% of the stars have a sub-solar metallicity. % and that more than half of the stars have a distance from the Galactic centre between 7 and 8~Kpc. 
For the LMC sample, we find a mean metallicity value of $\sim-0.33$~dex ($\sigma$ = 0.13), with a range of values between $-$0.62~dex and $-$0.10~dex. Here, the more metal-rich extreme is just an isolated case, while the metal-poor end of the distribution is represented by 3 stars.
For the SMC sample, we find that the mean value is about $\sim-0.75$ dex ($\sigma$ = 0.08), with a range of values between $-$0.87 and $-$0.63. While the average values for each galaxy are in good agreement with non-pulsating stars of similar age, Cepheids display a significant spread. Thus it is fundamental to measure the metallicity of individual stars.
Detailed discussion of these results and their comparison with previous studies will be presented in a forthcoming paper \citep{mot06}, along with an analysis of the effects of the metallicity on the Cepheid Period-Luminosity relation, also showed by Romaniello et al. in these proceedings.
In the very near future we will focus on the analysis of the $\alpha$-elements results and their interpretation.

\bibliographystyle{aa}

\begin{thebibliography}{}

\bibitem[Andrievsky et~al.(2002)]{and02} 
Andrievsky, S.M. et al.\ 2002, A\&A, 381, 32

\bibitem[Baraffe \& Alibert(2001)]{bar01} Baraffe, I. \& Alibert, Y.\  2001, A\&A, 371, 592

\bibitem[Bono et~al.(1999)]{bono99} Bono, G. et al.\ 1999, ApJ, 512, 711


\bibitem[Castelli \& Kurucz(2003)]{cas03} Castelli, F. \& Kurucz, R.L.\ 2003, in Modelling of Stellar Atmospheres, 210th IAU Symposium, Edited by N. Piskunov, W.W. Weiss and D.F. Gray. Published by the Astronomical Society of the Pacific, p.A20

\bibitem[Cayrel(1988)]{cay88} Cayrel, R.\ 1988, in The Impact of Very High S/N Spectroscopy on Stellar Physics, 132th IAU Symposium, Kluwer Academic Publishers, Dordrecht., p.345

\bibitem[Charbonneau(1995)]{char95} Charbonneau, P.\ 1995, ApJSS, 101, 309

\bibitem[Chiosi et~al.(1992)]{chiosi92} Chiosi, C. et al.\ 1992, ApJ, 387, 320 

\bibitem[Clementini et~al.(1995)]{clem95} Clementini, G. et al.\ 1995, AJ, 110, 2319

\bibitem[Fiorentino et~al.(2002)]{fio02} Fiorentino, G. et al.\ 2002, ApJ, 576, 402

\bibitem[Fry \& Carney(1997)]{fc97} Fry, A.M. \& Carney, B.W.\ 1997, AJ, 113, 1073

\bibitem[Groenewegen et~al.(2004)]{groe04} Groenewegen, M.A.T. et al.\ 2004, A\&A, 420, 655

\bibitem[Hill(1997)]{hill97} Hill, V.\ 1997, A\&A, 324, 435

\bibitem[Hill et~al.(1995)]{hill95} Hill, V., Andrievsky, S.M. \& Spite, M.\ 1995, A\&A, 293, 347

\bibitem[Kennicutt et~al.(1998)]{ken98} Kennicutt, R.C. et al.\ 1998, ApJ, 498, 181

\bibitem[Kiss \& Vinko(2000)]{kv00} Kiss, L.L. \& Vinko, J.\ 2000, MNRAS, 314, 420

\bibitem[Kovtyukh \& Gorlova(2000)]{kg00} Kovtyukh, V.V. \& Gorlova, N.I.\ 2000, A\&A, 351, 597

\bibitem[Kupka et~al.(1999)]{kup99} Kupka, F. et al.\ 1999, A\&AS, 138, 119

\bibitem[Kurucz(1993)]{ku93} Kurucz, R.L.\ 1993, CD-ROMS \#1, 13, 18

\bibitem[Laney \& Stobie(1994)]{ls94} Laney, C.D. \& Stobie, R.S.\ 1994, MNRAS, 226, 441 

\bibitem[Luck et~al.(1998)]{lu98} Luck, R.E. et al.\ 1998, AJ, 115, 605

\bibitem[Luck et~al.(2003)]{lu03} Luck, R.E. et al.\ 2003, A\&A, 401, 939

\bibitem[Marconi et~al.(2005)]{mar05} Marconi, M., Musella, I. \& Fiorentino, G.\ 2005, ApJ, 632, 590 

\bibitem[Mottini et~al.(2006)]{mot06} Mottini, M. et al.\ 2006, to be submitted to A\&A

\bibitem[Romaniello et~al.(2005)]{paper1} Romaniello, M. et al.\ 2005, A\&A, 429, L37 

\bibitem[Ryabchikova et~al.(1999)]{ryab99} Ryabchikova T.A. et al.\ 1999, proc. of the 6th International Colloquium on Atomic Spectra and Oscillator Strengths, Victoria BC, Canada, 1998, Physica Scripta T83, 162

\bibitem[Sakai et~al.(2004)]{sak04} Sakai, S. et al.\ 2004, ApJ, 608, 42

\bibitem[Sandage et~al.(1999)]{san99} Sandage, A., Bell, R.A. \& Tripicco, M.J.\ 1999, ApJ, 522, 250

\bibitem[Sasselov et~al.(1997)]{sas97} Sasselov, D.D. et al.\ 1997, A\&A, 324, 471

\bibitem[Storm et~al.(2004)]{storm04} Storm, J. et al.\ 2004, A\&A, 415, 531


\end{thebibliography}

\end{document}